\documentstyle[12pt]{report}
\def \a {\alpha}
\def \b {\beta}
\def \ep{\varepsilon}
\def \p {\partial}
\def \d {\mbox{D}}
\def \ts {\textstyle}
\def \be {\begin{equation}}
\def \ee{\end{equation}}
\def \bea{\begin{eqnarray}}
\def \eea{\end{eqnarray}}
\def \l {\label}
\def \m {\mbox}
\def \r {\ref}
\def \bi {\bibitem}
\def \ds {\displaystyle{\cdot}}
\begin{document}

\begin{titlepage}

\thispagestyle{empty}

\begin{center}
\[ \]
\[ \]
{\huge\bf CAUSAL THERMODYNAMICS \\

IN RELATIVITY\\

}
\[ \]
\[ \]
\[ \]
{\sf Lectures given at the\\

Hanno Rund Workshop on Relativity and Thermodynamics\\ 

Natal University, South Africa, June 1996}
\[ \]
\[ \]
\[ \]
\[ \]
Roy Maartens
\[ \]
{\em Department of Mathematics, Natal University, South Africa\\
and School of Mathematical Studies, Portsmouth University, PO1 2EG,
England} $^*$

\end{center}

\vfill
$^*${\footnotesize Permanent address. Email: maartens@sms.port.ac.uk}

\end{titlepage}

\tableofcontents

\chapter{Relativistic Fluid Dynamics}

In this chapter I review the basic theory of fluid kinematics
and dynamics (without dissipation) in relativistic spacetime. The
classic paper in this field is Ellis' 1971 review \cite{e}.
That paper is
at a more advanced level than these lectures.
For a basic
introduction to tensors, relativity and fluids,
see for example \cite{d}.

I use units in which the speed of light in vacuum, Einstein's
gravitational constant and Boltzmann's constant are all one:
$$
c=8\pi G=k=1
$$
I use $A\doteq B$ to denote equality of $A$ and $B$ in an
instantaneous orthonormal frame at a point (defined below).

\section{Brief Review of Relativity}

The observed universe is a 4 dimensional spacetime. Physical laws
should be expressible as equations in spacetime that are independent
of the observer. Together with experimental and observational
evidence, and further principles, this leads to Einstein's
relativity theory - special
relativity in the case where the gravitational field may be neglected,
and general relativity when gravity is incorporated.

Local coordinates, which are typically based on observers,
are usually chosen
so that $x^0$ is a time parameter and $x^i$ are space coordinates.
A change of coordinates (or of observers) is
\be
x^\a=(x^0,x^i)=(t,\vec{x})\quad\rightarrow\quad
x^{\a'}=(x^{0'},x^{i'})=(t',\vec{x}\,')
\l{1}\ee
Physical laws should then be invariant under such transformations.
This means that these laws are expressible in terms of tensor fields
and
tensor--derivatives. Tensors have different types $(r,s)$, but they
all transform linearly under (\r{1}). The simplest example is a 
scalar, which is invariant. Using the chain rule, the tranformation
of the coordinate differentials is seen to be linear:
$$
dx^{\a'}=\sum_\a{\p x^{\a'}\over\p x^\a}dx^\a\equiv {\p x^{\a'}\over
\p x^\a}dx^\a
$$
Extending this to partial derivatives of scalars and generalising,
we are led to the transformation properties of tensors in general:
\bea
(0,0)~\m{scalar} & & f~\rightarrow~ f \nonumber\\
(1,0)~\m{vector}  & & u^\a~\rightarrow~u^{\a'}={\p x^{\a'}\over
\p x^\a}u^{\a}
\nonumber\\
(0,1)~\m{covector} & & k_\a~\rightarrow~k_{\a'}={\p x^\a \over\p
x^{\a'}}k_\a \nonumber\\
(1,1)~\m{tensor}  & & T^\a{}_\b~\rightarrow~T^{\a'}{}_{\b'}
={\p x^{\a'}\over\p x^\a}{\p x^\b\over\p x^{\b'}}T^\a{}_\b
\nonumber\\
\cdots & & \cdots \nonumber\\
(r,s)~\m{tensor}  & & J^{\a_1\cdots\a_r}{}{}{}_{\b_1
\cdots\b_s}
~\rightarrow\nonumber\\
{}&& J^{\a_1'\cdots\a_r'}{}{}{}_{\b_1'\cdots\b_s'}=
{\p x^{\a_1'}\over\p x^{\a_1}}\cdots{\p x^{\b_s}\over\p x^{\b_s'}}
J^{\a_1\cdots\a_r}{}{}{}_{\b_1\cdots\b_s} \l{2}
\eea
It follows that if a tensor vanishes in one coordinate frame, it
vanishes in all frames. Consequently, if two tensors are equal in
one frame, they are equal in all frames.

Fields and equations that transform according to (\r{2}) are
called tensorial or covariant. Restricted covariance arises when
the class of allowable coordinate systems is restricted. In special
relativity (flat spacetime), one can choose orthonormal coordinates
$x^\a$ which correspond to inertial observers,
and if $x^{\a'}$ is required to be also orthonormal, then
\be
{\p x^{\a'}\over\p x^\a}=\Lambda^{\a'}{}_\a\quad\Leftrightarrow\quad
x^{\a'}=\Lambda^{\a'}{}_\a x^\a+C^\a
\l{3}\ee
where $\Lambda, C$ are constants and $\Lambda$ is a Lorentz matrix.
In other words, special relativity says that the laws of
physics (leaving aside gravity) are invariant under Lorentz
transformations that connect any inertial
observers in relative motion.
Under this restriction, the partial derivatives of tensors transform
according to (\r{2}), i.e. they are Lorentz covariant. We use the
notation
\be
J^{\a\cdots}{}{}_{\cdots\b,\mu}\equiv \p_\mu J^{\a\cdots}{}{}_
{\cdots\b}\equiv{\p\over\p x^\mu}J^{\a\cdots}{}{}_{\cdots\b}
\l{4}\ee
for partial derivatives. Thus in special relativity, physical laws
are expressed in orthonormal coordinates as PDE's; for example
the Klein--Gordon equation for a massless scalar field is
\be
\Box\Psi\equiv \eta^{\a\b}\p_\a\p_\b\Psi=0
\l{5}\ee
where
\be
\eta_{\a\b}=\m{diag}\,(-1,1,1,1)=\eta^{\a\b}
\l{6}\ee
are the orthonormal components of the metric tensor.

The metric $g_{\a\b}$
of any (in general curved) spacetime
determines the spacetime interval between events, the scalar
product of vectors, and the raising and lowering of indices on
general tensors:
\bea
ds^2 &=& g_{\a\b}dx^\a dx^\b \l{7}\\
u\cdot v &=& g_{\a\b}u^\a v^\b=u^\a v_\a=u_\a v^\a \l{8}\\
J^\a{}_{\b\mu} &=& g^{\a\nu}g_{\b\sigma}J_\nu{}^\sigma{}_\mu~,~~
\m{etc.} \l{9}
\eea
where the inverse metric is defined by
$g^{\a\mu}g_{\mu\b}=\delta^\a{}_\b$\,.

The metric is a symmetric tensor. For any rank--2 tensor, we can
define its covariant symmetric and skew parts:
\be
V_{(\a\b)}={\ts{1\over2}}\left(V_{\a\b}+V_{\b\a}\right)\,,\quad
V_{[\a\b]}={\ts{1\over2}}\left(V_{\a\b}-V_{\b\a}\right)
\l{9b}\ee
so that $g_{\a\b}=g_{(\a\b)}$.

At any point (or event) $P$,
an observer can choose coordinates $x^\a$ that bring
$g_{\a\b}(P)$ into orthonormal form. I will call such a
coordinate system an instantaneous orthonormal frame (IOF),
characterised by
\be
g_{\a\b}\doteq\eta_{\a\b}\quad\Leftrightarrow\quad
g_{\a\b}(P)\Big|_{\m{iof}}=\eta_{\a\b}
\l{9a}\ee
At each event along the
observer's worldline, the IOF is in general different.
In fact an IOF is orthonormal
in a neighbourhood of the original point $P$
if and only if the spacetime is locally flat.

In curved spacetime, the partial derivative (\r{3}) is not
covariant (except when $J$ is a scalar).
The metric defines a connection that `corrects' for
the variations in the coordinate basis (equivalently, that
provides a rule for parallel transport of vectors):
\be
\Gamma^\a{}_{\b\sigma}={\ts{1\over2}}g^{\a\mu}\left(
g_{\mu\b,\sigma}+g_{\sigma\mu,\b}-g_{\b\sigma,\mu}\right)=
\Gamma^\a{}_{(\b\sigma)}
\l{10}\ee
The connection, which is not a tensor since it corrects for
non--tensorial variations, defines a covariant derivative
\bea
f_{;\a} &=& f_{,\a} \nonumber\\
u^\a{}_{;\b} &=& u^\a{}_{,\b}+\Gamma^\a{}_{\mu\b}u^\mu
\nonumber\\
k_{\a;\b} &=& k_{\a,\b}-\Gamma^\mu{}_{\a\b}k_\mu \nonumber\\
\cdots & & \cdots \nonumber\\
J^{\a\cdots}{}{}_{\cdots\b;\sigma} &=& J^{\a\cdots}{}{}_{\cdots
\b,\sigma}+\Gamma^\a{}_{\mu\sigma}J^{\mu\cdots}{}{}_{\cdots\b}
+\cdots -\cdots -\Gamma^\mu{}_{\b\sigma}J^{\a\cdots}{}{}
_{\cdots\mu}
\l{11}\eea
We also write $\nabla_\sigma J^{\a\cdots}{}{}_{\cdots\b}$ for
the covariant derivative. One can always
find an IOF at any event $P$ such that the connection vanishes
at $P$:
\be
\Gamma^\a{}_{\b\gamma}\doteq 0\quad\Rightarrow\quad
J^{\a\cdots}{}{}_{\cdots\b;\mu}\doteq J^{\a\cdots}{}{}_{\cdots\b,\mu}
\l{11a}\ee
From now on, any IOF will be assumed to have this property.

The connection also defines a covariant
measure of spacetime curvature -- the Riemann tensor:
\be
R^\a{}_{\b\mu\nu}=-\Gamma^\a{}_{\b\mu,\nu}+\Gamma^\a{}
_{\b\nu,\mu}+\Gamma^\a{}_{\sigma\mu} \Gamma^\sigma{}_{\b\nu}
-\Gamma^\a{}_{\sigma\nu} \Gamma^\sigma{}_{\b\mu}
\l{12}\ee
Curvature is fundamentally reflected in the non--commutation
of covariant derivatives\footnote{except for scalars: 
$f_{;[\a\b]}=0$.}, as given by the Ricci identity
\be
u_{\a;\b\gamma}-u_{\a;\gamma\b}=R^\mu{}_{\a\b\gamma}u_\mu
\l{13}\ee
and its generalisations for higher rank tensors.
The trace--free part of the Riemann tensor is the Weyl tensor
$C^\a{}_{\b\mu\nu}$, which represents the `free' gravitational
field and describes gravity waves,
while the trace gives the Ricci tensor and Ricci scalar
\be
R_{\a\b}=R^\mu{}_{\a\mu\b}=R_{\b\a}\,,\quad R=R^\a{}_\a
\l{14}\ee
which are determined by the mass--energy--momentum distribution via
Einstein's field equations
\be
R_{\a\b}-{\ts{1\over2}}Rg_{\a\b}=T_{\a\b}
\l{15}\ee
where $T_{\a\b}$ is the energy--momentum tensor, discussed
below. The Ricci tensor obeys the contracted Bianchi identity
\be
\left(R^{\a\b}-{\ts{1\over2}}Rg^{\a\b}\right)_{;\b}=0
\l{15a}\ee

\section{Fluid Kinematics}

Consider the motion of a particle with rest
mass $m$. An observer records
the particle's history
-- its worldline -- as $x^\a=(t,x^i(t))$. We need a
covariant (observer--independent) description of the worldline
and velocity of the particle. If $m>0$, then along the
worldline $ds^2<0$ (the particle moves slower than light). If
$\tau$ is the time recorded by a clock comoving with the
particle, the worldline is
given by $x^\a=x^\a(\tau)$, independently of any
observer. The covariant comoving time is called the proper time.
In an IOF
$ds^2\doteq-d\tau^2$. Since both sides of this equation are tensors
(scalars), the equation holds in any frame, and at all
points along the worldline, i.e. $ds^2=-d\tau^2$.
The kinematics of the particle are covariantly described by
the 4--velocity
\be
u^\a={dx^\a\over d\tau}\quad\Rightarrow\quad u^\a u_\a =-1
\l{16}\ee
and the 4--acceleration
\be
\dot{u}^\a=u^\a{}_{;\b}u^\b
\l{17}\ee
where $\dot{u}^\a u_\a=0$.
The particle moves in free--fall, subject to no non--gravitational
forces, if and only if $\dot{u}_\a=0$, in which case its worldline is
a (timelike) geodesic.
In the observer's IOF
\be
u^\a\doteq\gamma(v)(1,{d\vec{x}\over dt})=\gamma(1,\vec{v})\,,\quad
\gamma(v)=(1-v^2)^{-1/2}={dt\over d\tau}
\l{18}\ee
where $t$ is the observer's proper time at that point, and
$\vec{v}$ is the measured velocity of the particle.

If $m=0$, the particle (photon or massless neutrino or graviton)
moves at the speed of
light, and along its worldline $ds^2=0$, so that proper time
cannot parametrise the worldline.
In the IOF of an observer $u^\a$, the light ray has angular
frequency $\omega$ and wave vector $\vec{k}$ (where $|\vec{k}|\doteq
\omega$), with phase $\phi\doteq\vec{k}\cdot\vec{x}-\omega t$, so
that
$$
\phi_{,\a}\doteq (-\omega,\vec{k})\quad\m{and}\quad\phi_{,\a}
\phi^{,\a}\doteq 0
$$
Now the phase is a covariant scalar, and its gradient is
a covariant null vector, which we call the 4--wave vector, and
which is geodesic:
\be
k_\a=\phi_{,\a}\quad\m{and}\quad k_\a k^\a=0\quad\Rightarrow\quad
k^\a{}_{;\b}k^\b=0
\l{19}\ee
From the above, in the observer's IOF, $\omega\doteq -k_\a u^\a
=\dot{\phi}$.
This gives a covariant expression for the redshift between
events $E$ (`emitter') and $R$ (`receiver') along a ray:
\be
1+z\equiv{\omega_E\over\omega_R}=
{\left(u_\a k^\a\right)_E\over\left(u_\a k^\a\right)_R}
\l{20}\ee
\\

A fluid is modelled as a continuum with a well--defined average
4--velocity field $u^\a$, where $u^\a u_\a=-1$.
This hydrodynamic description requires
that the mean collision time  is much less than any macroscopic
characteristic time (such as the expansion time in an expanding
universe); equivalently, the mean free path must be much less than
any macroscopic characteristic length. For a perfect fluid, $u^\a$
is uniquely defined\footnote{If the fluid is out of equilibrium as a
result of dissipative effects, then there is no unique average
4--velocity}
as the 4--velocity relative to which there is no
particle current, i.e.
\be
n^\a=n u^\a
\l{21}\ee
where $n$ is the number density.

The field of comoving observers $u^\a$ defines a covariant
splitting of spacetime into time $+$ space ($1+3$) via the
projection tensor
\bea
h_{\a\b}=g_{\a\b}+u_\a u_\b=h_{\b\a}~\Rightarrow &&
h_{\a\b}u^\b=0\,,~h_\a{}^\mu h_{\mu\b}=h_{\a\b} \nonumber\\
{}&& h^\a{}_\a=3\,,~
h_{\a\b}q^\b=q_\a~~\m{if}~~q_\a u^\a=0
\l{22}\eea
which projects at each point into the instantaneous rest space of
the fluid/ observer, and provides a 3--metric in the rest space.
In the comoving IOF
$$u^\a\doteq(1,\vec{0})\,,\quad h_{\a\b}\doteq\m{diag}~(0,1,1,1)\,,
\quad h_{\a\b}q^\a q^\b\doteq\vec{q}\cdot\vec{q}$$
where $q_\a u^\a=0$.
This allows us to compare relativistic fluid kinematics and
dynamics with its Newtonian limit.

The covariant time derivative along $u^\a$ is
\be
\dot{A}^{\a\cdots}{}{}_{\b\cdots}=A^{\a\cdots}{}{}_{\b\cdots;\mu}u^\mu
\l{23}\ee
and describes the rate--of--change relative to comoving observers. In
the comoving IOF
$$
\dot{A}^{\a\cdots}{}{}_{\b\cdots}\doteq{d\over d\tau}A^{\a\cdots}
{}{}_{\b\cdots}
$$
The covariant spatial derivative is
\bea
\d_\a f &=& h_\a{}^\b f_{,\b} \l{24}\\
\d_\a q_\b &=& h_\a{}^\mu h_\b{}^\nu \nabla_\mu q_\nu \l{25}\\
\d_\a\sigma_{\b\gamma} &=& h_\a{}^\mu h_\b{}^\nu h_\gamma{}^\kappa
\nabla_\mu\sigma_{\nu\kappa}\,,~~\m{etc.}
\l{26}\eea
and describes spatial variations relative to
comoving observers.
In the comoving IOF, with $q^\a\doteq(q^0,\vec{q}\,)$
\be
\d_\a f\doteq (0, \vec{\nabla}f)\,,\quad \d^\a q_\a\doteq
\vec{\nabla}\cdot\vec{q}\,,\quad \ep^{ijk}\d_j q_k\doteq \left(
\vec{\nabla}\times\vec{q}\right)^i
\l{27}\ee
\\

Any spacetime vector can be covariantly split as
\be
V^\a=Au^\a+B^\a\,,\quad\m{where}\quad A=-u_\a V^\a\,,
~B^\a=h^\a{}_\b V^\b~\Leftrightarrow~B^\a u_\a=0
\l{28}\ee
For a rank--2 tensor:
\be
V_{\a\b}=Au_\a u_\b+B_\a u_\b +u_\a C_\b+F_{\a\b}
\l{29}\ee
where $A=V_{\a\b}u^\a u^\b$, $B_\a u^\a=0= C_\a u^\a$ and
$$
F_{\a\b}=h_\a{}^\mu h_\b{}^\nu V_{\mu\nu}\quad\Leftrightarrow\quad
F_{\a\b}u^\a=0=F_{\a\b}u^\b
$$
For example, if $V_{\a\b}=W_{\a;\b}$, then $F_{\a\b}=\d_\b W_\a$.
Now $F_{\a\b}$ may be further decomposed into symmetric and
skew parts:
$$
F_{\a\b}=F_{(\a\b)}+F_{[\a\b]}
$$
In the comoving IOF, the corresponding decomposition of the
matrix of components $F_{ij}$ is simply
$$
F\doteq(F)+[F]={\ts{1\over2}}\left(F+F^T\right)
+{\ts{1\over2}}\left(F-F^T\right)
$$
and $(F)$ may be further split into its trace and
trace--free parts:
$$
(F)\doteq\left\{{\ts{1\over3}}
\m{tr}~F\right\}I+ \langle F\rangle
$$
The covariant expression of this is
$$
F_{(\a\b)}=\left\{{\ts{1\over3}}F^\gamma{}_\gamma
\right\}h_{\a\b}+ F_{<\a\b>}
$$
where the symmetric, spatial trace--free part of any tensor
is defined by
\be
V_{<\a\b>}=h_\a{}^\mu h_\b{}^\nu\left\{ V_{(\mu\nu)}-{\ts{1\over3}}
V_{\sigma\kappa}h^{\sigma\kappa}h_{\mu\nu}\right\}
\l{30}\ee
Thus we can rewrite the decomposition (\r{29}) in the covariant
irreducible form
\be
V_{\a\b}=Au_\mu u_\nu+B_\a u_\b+
u_\a C_\b+{\ts{1\over3}}V_{\mu\nu}h^{\mu\nu}h_{\a\b}
+V_{<\a\b>}+V_{[\mu\nu]}h^\mu{}_\a h^\nu{}_\b
\l{31}\ee
\\

Now we are ready to define the quantities that covariantly describe
the fluid kinematics. These quantities are simply the irreducible
parts of the covariant derivative of the fluid 4--velocity. With
$V_{\a\b}=u_{\a;\b}$, we have $A=0=C_\a$ since $u_{\a;\b}u^\a=0$,
and then $B_\a=-u_{\a;\b}u^\b=-\dot{u}_\a$. Thus (\r{31}) gives
\bea
&& u_{\a;\b}=Hh_{\a\b}+\sigma_{\a\b}+\omega_{\a\b}-\dot{u}_\a u_\b
\quad\m{where}\quad
3H=u^\a{}_{;\a}=\d^\a u_\a\,, \nonumber\\
&&\sigma_{\a\b}=
u_{<\a;\b>}=\d_{<\b}u_{\a>}\,,~~\omega_{\a\b}=h_\a{}^\mu h_\b{}^\nu
u_{[\mu;\nu]}=\d_{[\b}u_{\a]}
\l{32}\eea
In a comoving IOF at a point $P$, $\vec{v}$ is zero at $P$, but its
derivatives are not, and we find using (\r{27}) that
$$
3H\doteq\vec{\nabla}\cdot\vec{v}\,,~~\ep^{ijk}\omega_{jk}\doteq-\left(
\vec{\nabla}\times\vec{v}\right)^i
$$
so that $H$ generalises the Newtonian expansion rate and
$\omega_{\a\b}$ generalises the Newtonian vorticity.
Similarly, it
can be seen that $\sigma_{\a\b}$ is the relativistic generalisation
of the Newtonian shear. These kinematic quantities therefore have
the same physical interpretation as in Newtonian fluids.
A small sphere
of fluid defined in the IOF of a comoving observer
at $t=0$, and then measured in the observer's IOF a short time
later, undergoes the following changes:
\begin{itemize}
\item
due to $H$, its volume
changes but not its spherical shape;
\item
due to $\sigma_{\a\b}$, its
volume is unchanged but its shape is distorted in a way defined by
the eignevectors (principal axes) of the shear;
\item
due to $\omega_{\a\b}$, its volume and shape are unchanged, but it is
rotated about the direction $\vec{\nabla}\times\vec{v}$.
\end{itemize}

The expansion rate defines a comoving scale factor $a$ that determines
completely the volume evolution:
\be
H={\dot{a}\over a}
\l{33}\ee

\section{Conservation Laws - Perfect Fluids}

Assuming there are no unbalanced creation/ annihilation processes,
particle number is conserved in the fluid. In an IOF, this is
expressed via the continuity equation
$$
{\p n\over \p t}+\vec{\nabla}\cdot\left(n\vec{v}\right)\doteq 0
$$
By (\ref{21}), the covariant form of particle conservation is
\be
n^\a{}_{;\a}=0\quad\Leftrightarrow\quad \dot{n}+3Hn=0 \quad
\Leftrightarrow\quad n a^3=\m{comoving const}
\l{34}\ee
where (\r{33}) was used to show that the comoving particle number
$N\propto na^3$ is constant.

A perfect fluid is described by its 4--velocity $u^\a$,
number density $n$, energy
(or mass--energy) density $\rho$, pressure $p$ and specific
entropy $S$. In a comoving IOF, the pressure is isotropic and
given by the Newtonian stress tensor $\tau_{ij}\equiv p\delta_{ij}$.
This can be
covariantly combined with the energy density into the symmetric
energy--momentum tensor\footnote{The form of the energy--momentum
tensor may be justified via relativistic kinetic theory}
\be
T_{\a\b}=\rho u_\a u_\b+ph_{\a\b}
\l{35}\ee
so that $T_{00}\doteq\rho=\,$energy density, 
$T_{ij}\doteq\tau_{ij}=\,$momentum density, $T_{0i}\doteq0$.
Just as the divergence of $n^\a$ produces a conservation law (\r{34}),
so too does the divergence of $T^{\a\b}$:
\bea
T^{\a\b}{}{}_{;\b}=0~~\Rightarrow && \dot{\rho}+3H(\rho+p)=0
\l{36}\\
{}&& (\rho+p)\dot{u}_\a+\d_\a p=0
\l{37}\eea
In a comoving IOF these become
$$
{\p\rho\over\p t}+(\rho+p)\vec{\nabla}\cdot\vec{v}\doteq 0\,,~~
(\rho+p){\p\vec{v}\over\p t}\doteq-\vec{\nabla}p
$$
so that (\r{36}) is an energy conservation equation, generalising
the mass conservation equation of Newtonian fluid theory, while
(\r{37}) is a momentum conservation equation, generalising the
Euler equation. (In relativity, the pressure contributes to the
effective energy density.)
The energy--momentum conservation equation
also follows from Einstein's field equations (\r{15}) and the
contracted Bianchi identity (\r{15a}). Equivalently, the
conservation equation ensures that the identity holds, i.e. that
this integrability condition of the field equations is satisfied.

Finally, the entropy is also conserved. In a comoving IOF, there is
no entropy flux, and the specific entropy $S$ is constant for each
fluid particle. The covariant expression of this statement
is
\be
S^\a{}_{;\a}=0\quad\m{where}\quad S^\a=Sn^\a\quad\Rightarrow\quad
\dot{S}=0
\l{38}\ee
where (\r{34}) was used. Note that $S$ is constant along fluid
particle worldlines, and not throughout the fluid in general. If
$S$ is the same constant on each worldline -- i.e. if
$\d_\a S=0$ as well as $\dot{S}=0$, so that $S_{,\a}=0$ --
then the fluid is called isentropic.

\section{Equilibrium Thermodynamics}

A perfect fluid is characterised by $(n^\a, S^\a, T^{\a\b})$,
or equivalently by $(n,\rho,p,S, u^\a)$, subject to the
conservation laws above. What are the further relations amongst the
thermodynamic scalars $n,\rho,p,S$ and $T$, the temperature?
Firstly, the temperature is defined via the Gibbs equation
\be
TdS=d\left({\rho\over n}\right)+pd\left({1\over n}\right)
\l{39}\ee
where $df=f_{,\a}dx^\a$.
Secondly, thermodynamical equations of state are needed in order
to close the system of equations. Equations of state are
dependent on the particular physical properties of the fluid, and
are deduced from microscopic physics (i.e. kinetic
theory and statistical mechanics), or from phenomenological
arguments. In fact, assuming the metric is known (and so leaving
aside Einstein's field equations),
there are 7 equations -- i.e. (\r{34}),
(\r{36}), (\r{37})$_i$, (\r{38}), (\r{39}) -- for 8 variables
-- i.e. $n,\rho,p,u_i,S,T$.
Thus a single scalar equation of state will
close the system.

The Gibbs equation shows that in general two of the
thermodynamical scalars are needed as independent variables.
For example, taking $n,\rho$ as independent, the remaining
thermodynamical scalars are $p(n,\rho),S(n,\rho),T(n,\rho)$,
and given any one of these, say $p=p(n,\rho)$,
the others will be determined. Often a barotropic equation of
state for the pressure is assumed, i.e. $p=p(\rho)$. By the
Gibbs equation, this implies $S$ is constant (see 
below), i.e. the fluid is isentropic. 

The adiabatic
speed of sound $c_s$ in a fluid is given in general by
\be
c_s^2=\left({\p p\over\p\rho}\right)_S
\l{39b}\ee
For a perfect fluid, this becomes
\be
c_s^2={\dot{p}\over\dot{\rho}}
\l{39a}\ee
as can be seen by choosing $\rho, S$ as independent variables,
and using the fact that $\dot{S}=0$:
$$
\dot{p}=\left({\p p\over\p\rho}\right)_{\!S}\dot{\rho}+
\left({\p p\over\p S}\right)_{\!\rho}\dot{S}
$$

The preceding considerations are phenomenological and mathematical.
If the fluid model is based on microscopic physics, further
conditions are imposed. For example, if the fluid
is a collision--dominated gas in equilibrium, then relativistic
kinetic theory, based essentially on imposing energy--momentum
conservation at a microscopic level, leads to stringent
conditions\footnote{Note that kinetic theory incorporates
assumptions about the interactions of particles, in particular that
the interactions are described by the Boltzmann collision integral.}.
If $m>0$ is the rest mass of the particles and
$$
\beta_\mu={\beta \over m}u_\mu\,,\quad \beta={m\over T}
$$
then the following conditions hold:\footnote{See \cite{i}. 
In standard units, $\beta=mc^2/kT$.}
\bea
&& \beta_{(\mu;\nu)} = 0 \l{40}\\
&& mn=c_0{K_2(\beta)\over\beta}\,,~~~ p = nT \l{41}\\
&& \rho=c_0\left[{K_1(\beta)\over\beta}+3{K_2(\beta)\over\beta^2}
\right] \l{42}
\eea
where $c_0$ is a constant and $K_n$ are modified Bessel functions
of the second kind. Furthermore, (\r{40}) shows
that $\beta_\mu$ is a Killing vector field, so that the spacetime
is stationary. In particular, (\r{32}) implies
\be
H=0\,,\quad\dot{u}_\a=-\d_\a\ln T\,,\quad\sigma_{\a\b}=0
\l{43}\ee
and then (\r{34}), (\r{36}) lead to
\be
\dot{n}=\dot{\rho}=\dot{p}=\dot{T}=0
\l{44}\ee
Thus if the perfect fluid is a relativistic Maxwell--Boltzmann gas in
equilibrium, severe restrictions are imposed not only on the
fluid dynamics but also on the spacetime geometry.

In the case of a gas of massless particles in collisional equilibrium,
the conditions are less severe:
\bea
&& \beta_{(\mu;\nu)}=-{\dot{T}\over T^2}g_{\mu\nu}~~\Rightarrow~~
H=-{\dot{T}\over T}\,,~\sigma_{\a\b}=0 \l{45}\\
&& n=b_0 T^3\,,~~~\rho=3p=3nT \l{46}
\eea
Thus $\beta_\mu$ is a conformal Killing
vector field, so that expansion is possible in equilibrium.

Kinetic theory shows that a purely phenomenological approach to fluid
thermodynamics holds potential problems in the form of hidden or
unknown consistency conditions that may be violated. Any
phenomenological model needs to be applied with caution.

The best motivated barotropic
perfect fluid model is that for incoherent
radiation or massless particles,
for which $p={1\over3}\rho$,
as in (\r{46}). The energy conservation equation (\r{36}) integrates,
on using (\r{33}):
\be
\rho=(\m{comoving const}) a^{-4}
\l{46a}\ee
Cold, non--relativistic matter is often modelled as pressure--free
`dust', so that
\be
p=0\quad\Rightarrow\quad \rho=(\m{comoving const})a^{-3}=mn
\l{46b}\ee
A kinetic theory motivation for the dust model arises from (\r{41}),
(\r{42}) in the limit $\beta\gg 1$:
\be
p=nT\,,~~~\rho\approx mn+{\ts{3\over2}}nT\quad\m{where}\quad
T\ll m
\l{46c}\ee
The energy density is $\rho\approx n(mc^2+\ep)$, where $mc^2$ is
the rest mass energy per particle, and $\ep={3\over2}kT$ is the
thermal energy per particle.
While (\r{46c}) is still reasonable at high temperatures
(e.g. for the electron, $m\approx 10^9$K, and (\r{46c}) should
be very accurate for $T$ up to about $10^6$K), the
exact limiting dust case is only reasonable at low temperatures,
when random velocities are negligible. Of course the hydrodynamic
description is no longer valid in this limit.

We can find the evolution of the temperature easily in the
case of radiation. Comparing (\r{46}) and (\r{46a}), we get
\be
\m{radiation:}\quad\quad T\propto {1\over a}
\l{46e}\ee
In the general case, the Gibbs equation (\r{39}) can be written as
$$
dS=-\left({\rho+p\over Tn^2}\right)dn+{1\over Tn}d\rho
$$
and the integrability condition
$$
{\p^2S\over \p T\p n}={\p^2S\over\p n\p T}
$$
becomes
\be
n{\p T\over\p n}+(\rho+p){\p T\over\p\rho}=T{\p p\over\p\rho}
\l{46h}\ee
Furthermore,
since $T=T(n,\rho)$, it follows on using number and energy
conservation (\r{34}) and (\r{36}) that
$$
\dot{T}=-3H\left[n{\p T\over\p n}+(\rho+p){\p T\over \p \rho}\right]
$$
and then (\r{46h}) implies
\be
{\dot{T}\over T}=-3H\left({\p p\over\p\rho}\right)_{\!n}
\l{46f}\ee
From the derivation of (\r{46f}), we see that it will hold identically
if the Gibbs integrability condition, number conservation and
energy conservation are satisfied.

This equation holds for any perfect fluid. For
non--relativistic matter (\r{46c}) gives
$$
p={\ts{2\over3}}(\rho-mn)
$$
so that (\r{46f}) implies:
\be
\m{non--relativistic matter:}\quad\quad T\propto {1\over a^2}
\l{46g}\ee
This shows that the mean particle speed decays like $a^{-1}$,
since the thermal energy per particle is $\ep\approx {3\over2}
kT\approx{1\over2}m\bar{v}^2$.
Strictly, the limiting case of dust has $T=0$, but if dust is
understood as negligible pressure and temperature rather than
exactly zero pressure, then (\r{46g}) holds.

Note that the Gibbs integrability condition shows explicitly
that one cannot independently specify equations of state for
the pressure and temperature. This is clearly illustrated in the 
barotropic case.
\[ \]
{\bf Barotropic Perfect Fluids}\\

With $\rho, p$ as the independent variables in 
the Gibbs equation (\r{39}) in the general perfect fluid case, 
we find:
\bea
{n^2T\over(\rho+p)}dS &=&-\left[{\p n\over\p\rho}d\rho+
{\p n\over\p p}dp\right]+{n\over(\rho+p)}d\rho \nonumber\\
{}&=&\left[{n\over\rho+p}-{\p n\over\p\rho}\right]d\rho-
{\p n\over\p p}dp \nonumber\\
{}&=&\left[{n\over\rho+p}-{\dot{n}\over\dot{\rho}}+{\dot{p}\over
\dot{\rho}}{\p n\over\p p}\right]d\rho-{\p n\over\p p}dp \nonumber\\
{}&=& {\dot{p}\over\dot{\rho}}{\p n\over\p p}-{\p n\over\p p}dp 
\nonumber
\eea
where we used 
the conservation equations (\r{34}) and (\r{36}). Thus, for
any perfect fluid
\be
n^2TdS=(\rho+p){\p n\over\p p}\left[{\dot{p}\over\dot{\rho}}d\rho-dp
\right]
\l{46k}\ee
Suppose now that the pressure is barotropic: $p=p(\rho)$. It follows
immediately from (\r{46k}) that $dS=0$, i.e. the fluid is isentropic.

The same conclusion follows in the case of barotropic temperature.
Choosing $\rho, T$ as the independent variables, we find
$$
n^2TdS=(\rho+p){\p n\over\p T}\left[{\dot{T}\over\dot{\rho}}d\rho-dT
\right]
$$
so that $T=T(\rho)$ implies $dS=0$.

If the pressure and temperature are 
barotropic, then the Gibbs integrability
condition (\r{46h}) strongly restricts the form of $T(\rho)$:
\be
p=p(\rho)\quad\m{and}\quad T=T(\rho)\quad\Rightarrow\quad
T\propto \exp\int {dp\over\rho(p)+p}
\l{46i}\ee

The radiation and dust models are cases of a linear barotropic
equation of state that is often used for convenience
\be
p=(\gamma-1)\rho\quad\Rightarrow\quad \rho=(\m{comoving const})
a^{-3\gamma}
\l{46d}\ee
By (\r{39a}), the speed of sound is $c_s=\sqrt{\gamma-1}$.
For fluids which have some basis in kinetic theory,
one can impose the restriction $1\leq\gamma\leq{4\over3}$. In
principle ${4\over3}<\gamma\leq 2$ still leads to an allowable
speed of sound ($\gamma=2$ is known as `stiff matter').
The false
vacuum of inflationary cosmology may be formally described by the
case $\gamma=0$.

If (\r{46d}) holds then the Gibbs integrability condition (\r{46h})
becomes
$$
n{\p T\over\p n}+\gamma\rho{\p T\over\p \rho}=(\gamma-1)T
$$
whose solution by the method of characteristics yields
\be
T=\rho^{(\gamma-1)/\gamma}F\left({\rho^{1/\gamma}\over n}\right)
\l{46l}\ee
where $F$ is an arbitrary function. By (\r{34}) and (\r{46d}),
$F$ is a comoving constant, i.e. $\dot{F}=0$.
If $T$ is also barotropic, then $F$ is constant and
we have a power--law form with fixed exponent for the temperature:
\be
T\propto \rho^{(\gamma-1)/\gamma}
\l{46j}\ee
The same result follows directly from (\r{46i}).

Note that (\r{46d}) and (\r{46j}) are 
consistent with the ideal gas law $p=nT$. For
dissipative fluids, this is no longer true.

\section{Example: Cosmological Fluids}

The Ricci identity (\r{13}) for the fluid 4--velocity,
appropriately projected and contracted, together with the
field equations (\r{15}), leads to an evolution equation for
the expansion rate
\be
3\dot{H}+3H^2-\dot{u}^\a{}_{;\a}+\sigma_{\a\b}\sigma^{\a\b}
-\omega_{\a\b}\omega^{\a\b}=-{\ts{1\over2}}(\rho+3p)
\l{47}\ee
known as Raychaudhuri's equation.

In the standard FRW cosmological models, the rest spaces
of comoving observers mesh together to form spacelike 3--surfaces
$\{t=\,\m{const}\}$, where $t$ is proper time for comoving observers.
Each comoving observer sees that there are no preferred
spatial directions - i.e. the cosmic 3--surfaces are spatially
isotropic and homogeneous. Thus for any covariant scalar $f$
and vector $v_\a$
$$
\d_\a f=0~~~\left[\Leftrightarrow f=f(t)\right]\,,\quad h_\a{}^\b
v_\b=0~~~\left[\Leftrightarrow v_\a=V(t)u_\a\right]
$$
and
$$
u_{\a;\b}\propto h_{\a\b}\quad\Leftrightarrow\quad
\dot{u}_\a=0\,,~~~\sigma_{\a\b}=0=\omega_{\a\b}
$$
Raychaudhuri's equation (\r{47}) reduces to
\be
3\dot{H}+3H^2=-{\ts{1\over2}}(\rho+3p)
\l{48}\ee
The momentum conservation equation (\r{37}) is identically
satisfied. Since $\rho=\rho(t)$, $p=p(t)$, it follows that
$p=p(\rho)$, i.e. one may assume a barotropic equation of
state (for a single--component fluid).
Then (\r{48}) and the energy conservation equation (\r{36}) are
coupled equations in the 2 variables $H, \rho$, and can be
solved for a given $p(\rho)$. However it is more convenient to
use the Friedmann equation, the $(0,0)$ field equation, which is
a first integral of the Raychaudhuri equation:
\be
H^2={\ts{1\over3}}\rho-{k\over a^2}
\l{49}\ee
where $a(t)$ is the scale factor defined by (\r{33}) and $k=0,\pm 1$
is the curvature index for the cosmic 3--surfaces, which by symmetry
are spaces of constant curvature.
In comoving spherical coordinates,
the FRW metric and 4--velocity are
\be
ds^2=-dt^2+a(t)^2\left[{dr^2\over 1-kr^2}+r^2d\Omega^2\right]\,,\quad
u^\a=\delta^\a{}_0
\l{50}\ee
where $d\Omega^2$ is the metric of the unit sphere.

The expansion of the universe ($H>0$) is confirmed by the
systematic redshift in electromagnetic radiation that reaches us
from distant galaxies. By (\r{20}) and (\r{50})
\be
1+z={a(t_R)\over a(t_E)}
\l{50a}\ee
showing that $a$ is increasing,
so that by (\r{46d}) $\rho$ is decreasing. The early universe is
very hot, as confirmed by the after--glow we observe in the form
of the cosmic microwave background radiation. The early universe
is modelled by a radiation fluid (\r{46a}), while
the late universe is cold and the dust model (\r{46b}) is
appropriate. The transition from radiation-- to matter--domination
requires a careful analysis, and has to deal with the interaction
between radiation and matter. This covers the recombination era
of the universe, and involves dissipative processes which I will
discuss later.

Leaving aside this transition (which occupies a very
short time in the evolution of the universe), the matter and
radiation are effectively non--interacting. In the super--hot
conditions of the early universe, matter particles are
ultra--relativistic and effectively massless, so that a radiation
fluid in equilibrium is a good approximation. In the late universe,
(\r{46a}) and (\r{46b}) show that the energy density of
radiation is negligible compared to that of matter, and the
dust model becomes reasonable. For the flat universe case ($k=0$),
(\r{46a}), (\r{46b}) and (\r{49}) lead to the solutions:
\be
\m{radiation:}~~~ a\propto t^{1/2}\,,\quad\quad\m{matter:}~~~
a\propto t^{2/3}
\l{51}\ee

Einstein's theory predicts that a radiation FRW universe will
begin at $t=0$ with infinite energy density and temperature.
However, for times less than the Planck time $t_P\approx
10^{-43}$ sec, quantum gravity effects are expected to become
dominant, and Einstein's theory will no longer hold. As yet, no
satisfactory quantum gravity theory has been developed, and models
of the very early universe are necessarily speculative.
One fairly successful model, which applies during
the semi--classical
period between the quantum era and the classical Einstein era,
is inflation. Inflationary models aim to answer some of the
problems that arise in the standard classical cosmology
(the `big bang' model).

In these models, the energy density of the universe is
dominated by a scalar field at around $10^{-34}$ ---
$10^{-32}$ sec. The pressure of the scalar field is negative,
which acts like an effective repulsive force,
leading to accelerated expansion, or inflation, during
which the scale factor $a$ increases by around $10^{30}$.
Although the scalar field is not a fluid, it has an energy--momentum
tensor of the perfect fluid form (\r{35}). The condition for
accelerated expansion is $\ddot{a}>0$, so that, by (\r{48})
\be
\m{inflation}\quad\Leftrightarrow\quad \ddot{a}>0\quad
\Leftrightarrow\quad p<-{\ts{1\over3}}\rho
\l{52}\ee
Particular forms of inflation are exponential inflation in
a flat FRW universe, for which
\be
a\propto \exp(H_I t)\quad\m{and}\quad p=-\rho
\l{53}\ee
and power--law inflation, for which $a\propto t^N$, $N>1$.

\chapter{Dissipative Relativistic Fluids}

Perfect fluids in equilibrium generate no entropy and no
`frictional' type heating, because their dynamics is
reversible and without dissipation. For many processes in
cosmology and astrophysics, a perfect fluid model is adequate.
However, real fluids behave irreversibly, and some processes in
cosmology and astrophysics cannot be understood except as
dissipative processes, requiring a relativistic theory of
dissipative fluids.

In order to model such processes, we need non--equilibrium
or irreversible thermodynamics. Perhaps the most satisfactory
approach to irreversible thermodynamics is via non--equilibrium
kinetic theory. However, this is very complicated, and I will take
instead a standard phenomenological approach, pointing out how
kinetic theory supports many of the results. A comprehensive, modern
and accessible discussion of irreversible thermodynamics is given in
\cite{jcl}. This text includes relativistic thermodynamics, but most
of the theory and applications are non--relativistic. A
relativistic, but more advanced, treatment may be found in
\cite{i} (see also \cite{is}, \cite{hl}).

Standard, or classical, irreversible thermodynamics was first
extended from Newtonian to relativistic fluids by Eckart in
1940. However, the Eckart theory, and a variation of it due
to Landau and Lifshitz in the 1950's, shares with its Newtonian
counterpart the problem that dissipative perturbations propagate
at infinite speeds. This non--causal feature
is unacceptable in a relativistic theory -- and worse still, the
equilibrium states in the theory are unstable.

The problem is rooted in the way that non--equilibrium states
are described -- i.e. via the local equilibrium variables alone.
Extended irreversible thermodynamics takes its name from the fact
that the set needed to describe
non--equilibrium states is extended to include the dissipative
variables. This feature leads to causal and stable behaviour under
a wide range of conditions.
A non--relativistic
extended theory was developed by Muller in the 1960's,
and independently a relativistic version was developed by Israel
and Stewart in the 1970's. The extended theory is also known as
causal thermodynamics, second--order thermodynamics (because the
entropy includes terms of second order in the dissipative
variables), and transient thermodynamics (because the theory
incorporates transient phenomena on the scale of the mean free
path/ time, outside the quasi--stationary regime of the classical
theory).

In this chapter I will give a simple introduction to these
features, leading up to a formulation of relativistic
causal
thermodynamics that can be used for applications in cosmology
and astrophysics.

\section{Basic Features of Irreversible\protect\\ Thermodynamics}

For a dissipative fluid, the particle 4--current will be taken
to be of the same form as (\r{34}). This corresponds to
choosing an average 4--velocity in which there is no particle
flux -- known as the particle frame.
At any event in spacetime, the thermodynamic state of the fluid is
close to a fictitious equilibrium state at that event\footnote{Note
that the
equilibrium states are different at different events, and therefore
not subject to differential conditions such as (\r{40}) -- (\r{46})}.
The local
equilibrium scalars are denoted $\bar{n}, \bar{\rho}, \bar{p},
\bar{S}, \bar{T}$, and the local equilibrium 4--velocity is
$\bar{u}^\mu$. In the particle frame, it is possible to choose
$\bar{u}^\mu$ such that the number and energy densities coincide
with the local equilibrium values, while the pressure in general
deviates from the local equilibrium pressure:
\be
n=\bar{n}\,,\quad\rho=\bar{\rho}\,,\quad p=\bar{p}+\Pi
\l{1'}\ee
where $\Pi=p-\bar{p}$ is the bulk viscous pressure. From now on I
will drop the bar on the equilibrium pressure and write $p+\Pi$ for
the effective non--equilibrium pressure:
$$
p_{\m{eff}}=p+\Pi\quad\quad(p\rightarrow p_{\m{eff}}\,,
\quad\bar{p}\rightarrow p)
$$

The form of the
energy--momentum tensor may be deduced from
the equilibrium form (\r{35}) and the
general covariant decomposition (\r{31}), given that
$T_{\a\b}$ is symmetric:
\be
T_{\a\b}=\rho u_\a u_\b+(p+\Pi)h_{\a\b}+q_\a u_\b+q_\b u_\a+\pi_{\a\b}
\l{2'}\ee
where
$$
q_\a u^\a=0\,,\quad\pi_{\a\b}=\pi_{<\a\b>}~\Rightarrow~
\pi_{\a\b}u^\b=\pi_{[\a\b]}=\pi^\a{}_\a=0
$$
In a comoving IOF, $q_\a\doteq(0,\vec{q})$ and $\pi_{\a\b}\doteq
\pi_{ij}\delta_\a{}^i\delta_\b{}^j$, so
that $\vec{q}$ is an
energy flux (due to heat flow in the particle frame) relative to the
particle frame, while $\pi_{ij}$ is the anisotropic
stress.

Both the standard and extended theories
impose conservation of particle number and
energy--momentum:
$$
n^\a{}_{;\a}=0\,,\quad\quad T^{\a\b}{}{}_{;\b}=0
$$
Particle number conservation leads to the same equation (\r{34}) that
holds in the equilibrium case. However the equilibrium energy and
momentum conservation equations (\r{36}) and (\r{37}) are changed by
the dissipative terms in (\r{2'}):
\bea
\dot{\rho}+3H(\rho+p+\Pi)+\d^\a q_\a+2\dot{u}_\a q^\a+
\sigma_{\a\b}\pi^{\a\b}&=&0 \l{3'}\\
(\rho+p+\Pi)\dot{u}_\a+\d_\a(p+\Pi)+\d^\b\pi_{\a\b}+
\dot{u}^\b\pi_{\a\b}  &&{}\nonumber\\
{}+h_\a{}^\b\dot{q}_\b
+\left(4Hh_{\a\b}+\sigma_{\a\b}+\omega_{\a\b}\right)q^\b&=&0 \l{4'}
\eea

In irreversible thermodynamics, the entropy is no longer
conserved, but grows, according to the second law of thermodynamics.
The rate of entropy production is given by the divergence
of the entropy 4--current, so that
the covariant form of the second law of thermodynamics is
\be
S^\a{}_{;\a}\geq 0
\l{5'}\ee
$S^\a$ no longer has the simple form in (\r{38}), but
has a dissipative term:
\be
S^\a=Snu^\a+{R^\a\over T}
\l{6'}\ee
where $S=\bar{S}$ and $T=\bar{T}$ are
still related via the Gibbs equation
(\r{39}).\footnote{In extended thermodynamics, this is
the Israel--Stewart approach.
An alternative 
approach is to extend the Gibbs equation by
including
dissipative terms, and to use a generalised temperature, specific
entropy and pressure. The two approaches agree near 
equilibrium \cite{gl}.}

The dissipative part $R^\a$ of $S^\a$ is assumed to be an algebraic
function (i.e. not containing derivatives) of $n^\a$ and $T^{\a\b}$,
that vanishes in equilibrium:
$$
R^\a=R^\a(n^\b,T^{\mu\nu})\quad\m{and}\quad \bar{R}^\a=0
$$
This assumption is part of the hydrodynamical description, in the
sense that non--equilibrium states are assumed to be adequately
specified by the hydrodynamical tensors $n^\a, T^{\a\b}$
alone.\footnote{In kinetic theory, this corresponds to truncating
the non--equilibrium distribution function -- via the
Grad 14--moment approximation \cite{is}.}
The standard and extended theories of irreversible thermodynamics
differ in the form of this function.

\section{Standard Irreversible Thermodynamics}

The standard
Eckart theory makes the simplest possible assumption about $R^\a$
-- i.e. that it is linear in the dissipative quantities. The only
such vector that can be algebraically constructed from $(\Pi,q_\a,
\pi_{\a\b})$ and $u^\a$ is
$$
f(n,\rho)\Pi u^\a+g(\rho,n)q^\a
$$
Now the entropy density $-u_\a S^\a$ should be a maximum in
equilibrium, i.e.
$$
\left[{\p\over\p\Pi}(-u_\a S^\a)\right]_{\m{eqm}}=0
$$
This implies $f=0$.
In a comoving IOF,
$q_\a/T\doteq(0,\vec{q}/T)$, which is the entropy flux due to heat
flow. Thus $g=1$ and (\r{6'}) becomes
\be
S^\a=Snu^\a+{q^\a\over T}
\l{7'}\ee
Using the Gibbs equation (\r{39}) and
the conservation equations (\r{34}) and (\r{3'}),
the divergence of (\r{7'}) becomes
\be
TS^\a{}_{;\a}=-\left[3H\Pi+\left(\d_\a\ln T+\dot{u}_\a\right)q^\a
+\sigma_{\a\b}\pi^{\a\b}\right]
\l{8'}\ee
Notice that the equilibrium conditions (\r{43}) from kinetic theory
lead to the vanishing of each factor multiplying
the dissipative terms on the
right, and therefore to $S^\a{}_{;\a}=0$.

From (\r{8'}), we see that the simplest way to satisfy (\r{5'})
is to impose the following linear relationships between the
thermodynamic `fluxes' $\Pi, q_\a, \pi_{\a\b}$ and the
corresponding thermodynamic `forces' $H, \dot{u}_\a+\d_\a\ln T,
\sigma_{\a\b}$:
\bea
\Pi &=& -3\zeta H \l{9'}\\
q_\a &=& -\lambda \left(\d_\a T+T\dot{u}_\a\right) \l{10'}\\
\pi_{\a\b} &=& -2\eta \sigma_{\a\b} \l{11'}
\eea
These are the constitutive equations for dissipative quantities in the
standard Eckart theory of relativistic irreversible thermodynamics.
They are relativistic generalisations of the corresponding Newtonian
laws:
\bea
\Pi &=& -3\zeta \vec{\nabla}\cdot\vec{v}\quad\quad\m{(Stokes)}
\nonumber\\
\vec{q} &=& -\lambda \vec{\nabla} T\quad\quad\m{(Fourier)} \nonumber\\
\pi_{ij} &=& -2\eta \sigma_{ij}\quad\quad\m{(Newton)}\nonumber
\eea
This is confirmed by using a comoving IOF in (\r{9'}) -- (\r{11'}) --
except that in the relativistic case, as discovered by Eckart,
there is an acceleration term in (\r{10'})
arising from the inertia of heat
energy. Effectively, a heat flux will arise from accelerated matter
even in the absence of a temperature gradient.

The Newtonian laws allow us to identify the thermodynamic
coefficients:
\begin{itemize}
\item
$\zeta(\rho,n)$ is the bulk viscosity
\item
$\lambda(\rho,n)$ is the thermal conductivity
\item
$\eta(\rho,n)$ is the shear viscosity
\end{itemize}

Given the linear constitutive equations (\r{9'}) -- (\r{11'}), the
entropy production rate (\r{8'}) becomes
\be
S^\a{}_{;\a}={\Pi^2\over\zeta T}+{q_\a q^\a\over\lambda T^2}+
{\pi_{\a\b}\pi^{\a\b}\over2\eta T}
\l{12'}\ee
which is guaranteed to be non--negative provided that
$$
\zeta\geq 0\,,\quad\lambda\geq0\,,\quad\eta\geq0
$$

Note that the Gibbs equation (\r{39}) together with number and
energy conservation (\r{34}) and (\r{3'}), leads to an evolution
equation for the entropy:
\be
Tn\dot{S}=-3H\Pi-q^\a{}_{;\a}-\dot{u}_\a q^\a-\sigma_{\a\b}\pi^{\a\b}
\l{12a'}\ee

Many, probably most, of the applications of irreversible
thermodynamics in relativity have used this Eckart theory.
However the algebraic nature of the Eckart constitutive equations
leads to severe problems. Qualitatively, it can be seen that if
a thermodynamic force is suddenly switched off, then the
corresponding thermodynamic flux instantaneously vanishes. This
indicates that a signal propagates through the fluid at infinite
speed, violating relativistic causality.\footnote{Even in the
Newtonian case, infinite signal speeds present a problem, since
physically we expect the signal speed to be limited by the
maximum molecular speed.}

\section{Simple Example: Heat Flow}

For a quantitative demonstration, consider the flow of heat
in a non--accelerating, non--expanding and vorticity--free
fluid in flat spacetime, where the comoving
IOF may be chosen as a global orthonormal frame.
In the non--relativistic regime the fluid energy density is
given by (\r{46c}), and then the energy conservation equation
(\r{3'}) gives
$$
{\ts{3\over2}}n{\p T\over\p t}=-\vec{\nabla}\cdot\vec{q}
$$
since $\p n/\p t=0$ by (\r{34}). The Eckart law (\r{10'}) reduces to
$$
\vec{q}=-\lambda\vec{\nabla}T
$$
Assuming that $\lambda$ is constant, these two equations lead to
\be
{\p T\over\p t}=\chi\nabla^2 T\quad\m{where}\quad\chi={2\lambda
\over 3n}
\l{13'}\ee
which is the heat conduction equation. This equation is parabolic,
corresponding to infinite speed of propagation.

Apart from causality violation, the Eckart theory has
in addition the pathology of unstable equilibrium states. It can
be argued that a dissipative fluid will very rapidly tend
towards a quasi--stationary state that is adequately described by
the Eckart theory. However, there are many processes in which
non--stationary relaxational effects dominate.\footnote{For 
examples and further discussion, see \cite{jcl}, \cite{s}.}
Furthermore, even
if the Eckart theory can describe the asymptotic states, it is
clearly unable to deal with the evolution towards these states,
or with the overall dynamics of
the fluid, in a satisfactory way.

Qualitatively, one expects that if a thermodynamic force is
switched off, the corresponding thermodynamic flux should die away
over a finite time period. Referring to the heat flow example above,
if $\vec{\nabla}T$ is set to zero at time $t=0$, then instead
of $\vec{q}(t)=0$ for $t\geq0$,
as predicted by the Eckart law, we expect that
$$
\vec{q}(t)=\vec{q}_0\exp\left(-{t\over\tau}\right)
$$
where $\tau$ is a characteristic relaxational time for transient heat
flow effects. Such a relaxational feature would arise if the
Eckart--Fourier law were modified as
\be
\tau\dot{\vec{q}}+\vec{q}=-\lambda\vec{\nabla}T
\l{14'}\ee
This is the Maxwell--Cattaneo modification of the Fourier law,
and it is
in fact qualitatively what arises in the extended theory.

With the Maxwell--Cattaneo form (\r{14'}), the heat conduction
equation (\r{13'}) is modified as
\be
\tau{\p^2T\over\p t^2}+{\p T\over\p t}-\chi\nabla^2 T=0
\l{13a'}\ee
which is a damped wave equation. A thermal plane--wave solution
$$
T\propto \exp\left[i(\vec{k}\cdot\vec{x}-\omega t)\right]
$$
leads to the dispersion relation
$$
k^2={\tau\omega^2\over\chi}+i\omega
$$
so that the phase velocity is
$$
V={\omega\over\m{Re}(k)}=\left[{2\chi\omega\over \tau\omega+
\sqrt{1+\tau^2\omega^2}}\right]^{1/2}
$$
In the high frequency limit, i.e. $\omega\gg \tau^{-1}$, we see that
$$
V\approx \sqrt{{\chi\over\tau}}
$$
The high--frequency limit gives the speed of thermal pulses --
known as second sound -- and
it follows that this speed is finite for $\tau>0$.
Thus the introduction of a
relaxational term removes the problem of infinite propagation
speeds.

The intuitive arguments of this section form an introduction to the
development of the extended theory of Israel and Stewart.

\section{Causal Thermodynamics}

Clearly the Eckart postulate (\r{7'}) for $R^\a$ is too simple.
Kinetic theory indicates that in fact $R^\a$ is second--order in the
dissipative fluxes. The Eckart assumption, by truncating at first
order, removes the terms that are necessary to provide causality and
stability. The most general algebraic form for $R^\a$ that is at
most second--order in the dissipative fluxes is
\bea
S^\mu  &=& Snu^\mu+{q^\mu\over T}-
\left(\beta_0\Pi^2
+\beta_1q_\nu q^\nu+\beta_2\pi_{\nu\kappa}
\pi^{\nu\kappa}\right){u^\mu\over 2T} \nonumber\\
{}&& +{\alpha_0\Pi q^\mu\over T}+{\alpha_1\pi^{\mu\nu}q_\nu\over T}
\l{15'}\eea
where $\beta_A(\rho,n)\geq0$  are thermodynamic
coefficients for scalar, vector and tensor dissipative contributions
to the entropy density, and
$\alpha_A(\rho,n)$ are thermodynamic
viscous/ heat coupling coefficients. It follows from (\r{15'}) that
the effective entropy density (measured by comoving observers) is
\be
-u_\mu S^\mu=Sn-
{1\over2T}\left(\beta_0\Pi^2
+\beta_1q_\mu q^\mu+\beta_2\pi_{\mu\nu}
\pi^{\mu\nu}\right)
\l{16'}\ee
independent of $\alpha_0, \alpha_1$.
(Note that the entropy density is a maximum in equilibrium.)

For simplicity, I will assume
\be
\alpha_0=0=\alpha_1\quad\quad\m{i.e. no viscous/ heat coupling}
\l{17'}\ee
This assumption is consistent with linearisation in a perturbed
FRW universe, since the coupling terms lead to non--linear
deviations from the FRW background. However, the assumption (\r{17'})
may not be reasonable for non--uniform stellar models and other
situations where the background solution is inhomogeneous.

The divergence of the extended current (\r{15'}) --
with (\r{17'}) -- follows
from the Gibbs equation and
the conservation equations (\r{34}), (\r{3'}) and (\r{4'}):
\bea
TS^\a{}_{;\a} &=& -\Pi\left[3H+\beta_0\dot{\Pi}+{\ts{1\over2}}
T\left({\beta_0\over T}u^\a\right)_{;\a}\Pi\right]\nonumber\\
{}&& -q^\a\left[\d_\a\ln T+\dot{u}_\a+\beta_1\dot{q}_\a+{\ts{1\over2}}
T\left({\beta_1\over T}u^\mu\right)_{;\mu}q_\a\right]\nonumber\\
{}&&-\pi^{\a\mu}\left[\sigma_{\a\mu}+\beta_2\dot{\pi}_{\a\mu}+
{\ts{1\over2}}T\left({\beta_2\over T}u^\nu\right)_{;\nu}
\pi_{\a\mu}\right]
\l{18'}\eea
The simplest way to satisfy the second law of thermodynamics (\r{5'}),
is to impose, as in the standard theory, linear relationships 
between the thermodynamical fluxes and forces (extended), leading to
the following constitutive or transport equations\footnote{This linear
assumption is in fact justified by kinetic theory, which leads to the
same form of the transport equations \cite{is}.}:
\bea
\tau_0\dot{\Pi}+\Pi &=& -3\zeta H-\left[{\ts{1\over2}}\zeta T
\left({\tau_0\over\zeta T}u^\a\right)_{;\a}\Pi\right]  \l{19'}\\
\tau_1 h_\a{}^\b\dot{q}_\b+q_\a &=& -\lambda\left(\d_\a T+T\dot{u}_\a
\right)-\left[{\ts{1\over2}}\lambda T^2\left({\tau_1\over\lambda T^2}
u^\b\right)_{;\b}q_\a\right] \l{20'}\\
\tau_2 h_\a{}^\mu h_\b{}^\nu\dot{\pi}_{\mu\nu}+\pi_{\a\b} &=&
-2\eta\sigma_{\a\b}-\left[\eta T\left({\tau_2\over 2\eta T}u^\nu
\right)_{;\nu}\pi_{\a\b}\right]  \l{21'}
\eea
where the relaxational times $\tau_A(\rho,n)$ are given by
\be
\tau_0=\zeta\beta_0\,,\quad\tau_1=\lambda T\beta_1\,,\quad
\tau_2=2\eta\beta_2
\l{21a'}\ee
With these transport equations, the entropy production rate
has the same non--negative form (\r{12'}) as in the standard theory.

Because of the simplifying assumption (\r{17'}), there are no
couplings of scalar/ vector/ tensor dissipative fluxes.
As well as these viscous/ heat couplings, kinetic theory shows
that in general there will also be couplings of heat flux and
anisotropic pressure to the vorticity -- which, unlike the shear,
does not vanish in general
in equilibrium (see (\r{43})). These couplings
give rise to the following additions to the right hand
sides of (\r{20'}) and (\r{21'}) respectively:
$$
+\lambda T\gamma_1 \omega_{\a\b}q^\b\quad\quad\m{and}\quad\quad
+2\eta\gamma_2 \pi^\mu{}_{<\a}\omega_{\b>\mu}
$$
where $\gamma_1(\rho,n),\gamma_2(\rho,n)$ are the thermodynamic
coupling coefficients. In a comoving IOF, (\r{27}) shows that the
addition to (\r{20'}) has the form
$$
\lambda T\gamma_1 \vec{\omega}\times\vec{q}\quad\m{where}\quad
\vec{\omega}\doteq \vec{\nabla}\times\vec{v}
$$
If the background solution has zero
vorticity, as is the case in a perturbed FRW universe, then these
vorticity coupling terms will vanish in linearised theory.
However, they would be important in rotating stellar models, where
the background equilibrium solution has $\omega_{\a\b}\neq0$.

The terms in square brackets on the right
of equations (\r{19'}) -- (\r{21'}) are often omitted. This
amounts to the implicit assumption that these terms are negligible
compared with the other terms in the equations. I will call
the simplified equations the truncated Israel--Stewart equations.
One needs to investigate carefully the conditions under which the
truncated equations are reasonable. This will
be further discussed in the next chapter. The truncated equations,
together with the no--coupling assumption (\r{17'}), are of
covariant relativistic
Maxwell--Cattaneo form:
\bea
\tau_0\dot{\Pi}+\Pi &=& -3\zeta H
\l{22'}\\
\tau_1 h_\a{}^\b\dot{q}_\b+q_\a &=& -\lambda\left(\d_\a T+T\dot{u}_\a
\right)
\l{23'}\\
\tau_2 h_\a{}^\mu h_\b{}^\nu\dot{\pi}_{\mu\nu}+\pi_{\a\b} &=&
-2\eta\sigma_{\a\b}
\l{24'}
\eea

The crucial difference between the standard Eckart and the extended
Israel--Stewart transport equations is that the latter are differential
evolution equations, while the former are algebraic relations. As we
saw in the previous section, the evolution terms, with the relaxational
time coefficients $\tau_A$, are needed for causality -- as well as
for modelling high--frequency or transient phenomena,
where `fast' variables and relaxation effects
are important. The price paid for the improvements that the
extended causal thermodynamics brings is
that new thermodynamic coefficients are introduced.
However, as is the case with the coefficients $\zeta,\lambda,\eta$
that occur also in standard theory, these new coefficients may be
evaluated or at least estimated via kinetic theory. The
relaxation times $\tau_A$
involve complicated collision integrals. In fact, they are
usually estimated as mean collision times,
of the form
\be
\tau\approx{1\over n\sigma v}
\l{25'}\ee
where $\sigma$ is a collision cross section and $v$ the mean
particle speed.

It is important to remember that the derivation of the causal
transport equations is based on the assumption that the fluid is
close to equilibrium. Thus the dissipative fluxes are small:
\be
|\Pi|\ll p\,,\quad \left(\pi_{\a\b}\pi^{\a\b}\right)^{1/2}
\ll p\,,\quad \left(q_\a q^\a\right)^{1/2}\ll \rho
\l{26'}\ee

Consider the evolution of entropy in the Israel--Stewart theory.
The equation (\r{12a'}) still holds in the extended case:
\be
Tn\dot{S}=-3H\Pi-q^\a{}_{;\a}-\dot{u}_\a q^\a-\sigma_{\a\b}\pi^{\a\b}
\l{27'}\ee
Consider a comoving volume of fluid, initially of size
$a_0^3$, where $a$ is the scale factor defined in general by
(\r{33}). The entropy in this comoving volume is given by
\be
\Sigma=a^3nS
\l{28'}\ee
Then, by virtue of number conservation (\r{34}) and (\r{27'}),
it follows that
the growth in comoving entropy over a proper time
interval $t_0\rightarrow t$ is
\be
\Sigma(t)=\Sigma_0-\int_{t_0}^t{a^3\over T}\left(3H\Pi+
q^\a{}_{;\a}+\dot{u}_\a q^\a+\sigma_{\a\b}\pi^{\a\b}\right)dt
\l{29'}\ee
The second law, which is built into the theory, guarantees that
$\Sigma(t)\geq\Sigma_0$. However, it is possible that the local
equilibrium specific entropy $S$ is {\em not} increasing at all 
times -- but the effective, non--equilibrium specific entropy
$-u_\a S^\a/n$ is monotonically increasing \cite{jcl}.

Next we look at the temperature behaviour
in causal thermodynamics. The Gibbs integrability condition (\r{46h})
still holds:
\be
n{\p T\over\p n}+(\rho+p){\p T\over\p\rho}=T{\p p\over\p\rho}
\l{30'}\ee
However, the change in the energy conservation equation (\r{3'})
leads to a generalisation of the temperature evolution (\r{46f}):
\bea
{\dot{T}\over T} &=& -3H\left({\p p\over\p\rho}\right)_{\!n}
-{1\over T}
\left({\p T\over\p\rho}\right)_{\!n}\left[3H\Pi+q^\a{}_{;\a}+
\dot{u}_\a q^\a+\sigma_{\a\b}\pi^{\a\b}\right]
\l{31'}\\
&=&-3H\left({\p p\over\p\rho}\right)_{\!n}+n\dot{S}\left({\p T\over
\p\rho}\right)_{\!n} \nonumber
\eea

Note that if the Gibbs integrability condition, number conservation
and energy conservation are satisfied, then the evolution
equation (\r{31'}) will be an identity. This evolution equation
shows quantitatively how the relation of temperature to
expansion is affected by dissipation. 
The first term on the right of (\r{31'}) represents the cooling due
to expansion. In the second, dissipative term, viscosity in general 
contributes to heating effects, while the contribution of heat flow 
depends on whether heat is being transported into or out of a
comoving volume.

If instead of $(n,\rho)$ we choose $(n,T)$ as independent variables,
then the Gibbs integrability condition (\r{30'}) becomes
\be
T{\p p\over\p T}+n{\p\rho\over\p n}=\rho+p
\l{32'}\ee
and the temperature evolution equation (\r{31'}) becomes
\bea
{\dot{T}\over T} &=& -3H\left({\p p/\p T \over\p\rho/\p T}
\right)_{\!n}-{1\over T
(\p \rho/\p T)_n}\left[3H\Pi+q^\a{}_{;\a}+\dot{u}_\a
q^\a+\sigma_{\a\b}\pi^{\a\b}\right]
\l{33'}\\
&=&-3H\left({\p p/\p T \over\p\rho/\p T}\right)_{\!n} +
n\dot{S}{1\over(\p\rho/\p T)_n} \nonumber
\eea

Finally, we consider briefly the issue of equations of state for the
pressure and temperature in dissipative fluids.  Using the energy
conservation equation (\r{3'}),
the Gibbs equation in the form (\r{46k}) generalises to
$$
n^2TdS = \left[{n{\cal D}
\over 3H(\rho+p)+{\cal D}}\right]
d\rho +(\rho+p){\p n\over\p p}\left[{\dot{p}\over\dot{\rho}}d\rho
-dp\right] 
$$
where the dissipative term is
\be
{\cal D}=3H\Pi+q^\a{}_{;\a}+\dot{u}_\a q^\a+
\sigma_{\a\b}\pi^{\a\b}
\l{34'}\ee
It follows that in the presence of dissipation, barotropic 
pressure no longer forces $dS$ to vanish:
\be
dS ={1\over nT} \left[{{\cal D}\over 3H(\rho+p)+{\cal D}}\right]d\rho
\l{35'}\ee

As in the equilibrium case, it remains true, via the Gibbs 
integrability condition, that barotropic $T=T(\rho)$ together with
$p=(\gamma-1)\rho$ leads to the power--law form (\r{46j}) for
the temperature. However, in the dissipative case, these relations
are not in general compatible with the ideal gas law $p=nT$:
\bea
p=nT\,,~~p=(\gamma-1)\rho\,,~~T\propto \rho^{(\gamma-1)/\gamma}
&\Rightarrow& n\propto \rho^{1/\gamma}\nonumber\\
{}&\Rightarrow & {\dot{n}\over n}={1\over\gamma}{\dot{\rho}\over\rho}
\nonumber 
\eea
Then number and energy conservation imply ${\cal D}=0$.

However, it is possible to impose the $\gamma$--law and the ideal gas
law simultaneously, provided the temperature is not barotropic.
The temperature evolution equation (\r{33'}) and energy conservation
(\r{3'}) give
\be
{\dot{T}\over T}=\left[\left({\gamma-1\over\gamma}\right)
{\dot{\rho}\over\rho}+{{\cal D}\over\gamma\rho}\right]\left[1+
{{\cal D}\over nT}\right]
\l{36'}\ee

These results have interesting implications for a dissipative fluid
which is close to a 
thermalised radiation fluid, i.e. $p={1\over3}\rho$. If we
insist that $p=nT$, then the Stefan--Boltzmann law $\rho\propto T^4$
cannot hold out of equilibrium. Alternatively, if we impose the
Stefan--Boltzmann law, then the ideal gas law cannot hold unless
the fluid returns to equilibrium.

\chapter{Applications to Cosmology and Astrophysics}

The evolution of the
universe contains a sequence of important dissipative
processes, including:
\begin{itemize}
\item GUT (Grand Unified Theory) phase transition ($t\approx
10^{-34}$ sec, $T\approx 10^{27}$~K),
when gauge bosons acquire mass (spontaneous symmetry breaking).
\item
Reheating of the universe at the end of inflation (at about
$10^{-32}$ sec), when the
scalar field decays into particles.
\item
Decoupling of neutrinos from the cosmic plasma
($t\approx 1$ sec, $T\approx 10^{10}$ K), when the temperature
falls below the threshold for interactions that keep the
neutrinos in thermal contact.
The growing neutrino mean free path leads to
heat and momentum transport by neutrinos and thus damping of
perturbations. Shortly after decoupling, electrons and positrons
annihilate, heating up the photons in a non--equilibrium process.
\item
Nucleosynthesis (formation of light nuclei) ($t\approx 100$ sec).
\item
Decoupling of photons from matter during the recombination era
($t\approx 10^{12}$ sec, $T\approx 10^3$ K), when electrons
combine with protons and so no longer scatter the photons.
The growing photon mean free path leads to heat and momentum
transport and thus damping.
\end{itemize}

\noindent Some astrophysical dissipative processes are:
\begin{itemize}
\item
Gravitational collapse of local inhomogeneities to form
galactic structure, when viscosity and heating lead to dissipation.
\item
Collapse of a radiating star to a neutron star or black hole,
when neutrino emission is responsible for dissipative heat
flow and viscosity.
\item
Accretion of matter around a neutron star or black hole.

\end{itemize}

Further discussion of such processes can be found in \cite{kt},
\cite{cl} (but not from a causal thermodynamics standpoint).
The application of causal thermodynamics to cosmology and astrophysics
remains relatively undeveloped -- partly because of the complexity
of the transport equations, partly because all of the important
dissipative processes have been throughly analysed using the
standard theory or kinetic theory or numerical methods.

Causal bulk viscosity in cosmology
has been fairly comprehensively investigated - see \cite{bnk} --
\cite{z2}. Shear viscosity in
anisotropic cosmologies has been considered in \cite{bnk},
\cite{rp}, while
heat flow in inhomogeneous cosmologies has been discussed in
\cite{tp}. Causal dissipation in astrophysics has been
investigated in \cite{s}, \cite{fl} -- \cite{mz}. 
In all of these papers,
it is found that causal thermodynamic 
effects can have a significant impact and can
lead to predictions very different from those in the standard
Eckart theory.

In this chapter I will briefly
discuss some overall features of causal thermodynamics in
a cosmological/ astrophysical setting, and then conclude with
a more detailed discussion of bulk viscosity in an FRW universe,
which is the most accessible problem.

\section{General Features of Cosmic Dissipation}

The expanding universe defines a natural time--scale  -- the
expansion time $H^{-1}=a/\dot{a}$. Any particle species will
remain in thermal equilibrium with the cosmic fluid so long as
the interaction rate is high enough to allow rapid adjustment to
the falling temperature. If the mean interaction time is $t_c$, then
a necessary condition for maintaining thermal equilibrium is
\be
t_c < H^{-1}
\l{1.}\ee
Now $t_c$ is determined by 
\be
t_c={1\over n\sigma v}
\l{2.}\ee
where $\sigma$ is
the interaction cross--section,
$n$ is the number density of the target particles with which
the given species is interacting, and $v$ is the mean relative
speed of interacting particles. 

As an example, consider neutrinos in the early universe. At high
enough temperatures, the neutrinos are kept in thermal equilibrium
with photons and electrons via interactions with electrons that
are governed by the weak interaction. The cross--section is
\be
\sigma_w=g_0 T^2
\l{3.}\ee
where $g_0$ is a constant. The number density of electrons is
$n\propto T^3$, by (\r{46}), since the electrons are effectively
massless at these very high temperatures. Since $v=1$, (\r{2.})
gives $t_c \propto T^{-5}$. By (\r{46e}), we can see that
$H\propto T^2$. Thus
\be
t_cH=\left({T_*\over T}\right)^3
\l{4.}\ee
and using (\r{1.}) and the numerical values of the various constants,
it follows that the neutrinos will decouple for
$$
T<T_*\approx 10^{10}\,\m{K}
$$

Other 
cosmic decoupling processes may be analysed by a similar approach.
The differences arise from the particular forms of
$\sigma(T)$, $n(T)$ and $H(T)$. For example,
in the case of photons interacting with electrons via Thompson
scattering, the Thompson cross--section is constant, while the
number density of free electrons is given by a complicated
equation (the Saha equation),
which takes account of the process of recombination. The
expansion rate $H$ is also fairly complicated, since the universe
is no longer radiation--dominated. One finds that the decoupling
temperature is about $10^3$ K.

In the case of a collapsing star, similar arguments are applied --
except that the characteristic time in this case is determined
by the rate of collapse, which is governed by stellar dynamics.
For example, for neutrinos in the core of a neutron star,
interactions with electrons and nucleons determine an interaction
time that must be compared with the collapse time to estimate the
decoupling conditions for
the neutrinos -- after which they transport heat and
momentum away from the core.

The entropy generated in a dissipative process that begins at
$t_0$ and ends at $t_0+\Delta t$ is given by (\r{29'}):
\be
\Delta\Sigma=-\int_{t_0}^{t_0+\Delta t}{a^3\over T}\left(3H\Pi+
q^\a{}_{;\a}+\dot{u}_\a q^\a+\sigma_{\a\b}\pi^{\a\b}\right)dt
\l{5.}\ee
For example, $\Delta t$ could be
the time taken for a decoupling process in the universe or
a star.

The observed universe
has a high entropy, as indicated by the high number of photons per
baryon, about $10^8$. This gives a total entropy in the observable
universe of about $10^{88}$. Inflationary cosmology predicts that
nearly all of this entropy is generated by the reheating process at
the end of inflation -- i.e. that all other dissipative processes
in the evolution of the universe make a negligible contribution to
entropy production by comparison. In this model, the formula
(\r{5.}) would have to be modified to include
the dissipation not just from the fluid effects
that we have been discussing, but also from particle production.
Particle production, at a rate $\nu$, leads to non--conservation
of particle number, so that (\r{34}) is replaced by
\be
n^\a{}_{;\a}=\dot{n}+3Hn=\nu n
\l{8.}\ee
Then it is found that $\nu$ contributes to entropy production.
The contribution from particle production may be modelled
as an effective bulk viscosity.

Many dissipative processes are well described
by a radiative fluid -- i.e. a fluid consisting of interacting
massless and massive particles. The radiative fluid is dissipative,
and kinetic theory or fluctuation theory arguments may be used to
derive the dissipative coefficients in terms of the relaxation
times $\tau_A$ (which are usually assumed equal to the appropriate
interaction time $t_c$). The results are collected in the table below.
The table also includes the case of a relativistic Maxwell--Boltzmann
gas -- i.e. a dilute monatomic gas with high collision rate -- in
both the ultra--relativistic and non--relativistic limits. The
local equilibrium energy density and pressure are given by the
equations of state (\r{41}), (\r{42}) (but not subject to the
global equilibrium conditions (\r{43}), (\r{44})).
\[ \]
\[ \]

$$
\begin{array}{|l|c|c|c|}  \hline
{}&{}&{}&{} \\
{}& \zeta & \lambda & \eta \\
{}&{}&{}&{} \\ \hline\hline
{}&{}&{}&{} \\
\m{radiative fluid}&{}&{}&{}\\
\m{(massless/ massive)}& 4r_0T^4\Gamma^2\tau_0 & {\ts{4\over3}}r_0
T^3c^2\tau_1 & {\ts{4\over15}}r_0T^4\tau_2 \\
{}&{}&{}&{} \\ \hline
{}&{}&{}&{} \\
\m{Maxwell--Boltzmann gas:}&{}&{}&{} \\
\m{ultra--relativistic~} (\beta\ll 1) & {\ts{1\over216}}\beta^4 p\tau_0
& {\ts{4\over5}}T^{-1}p\tau_1 & {\ts{2\over3}}p\tau_2 \\
{}&{}&{}&{}
\\ \hline
{}&{}&{}&{} \\
\m{Maxwell--Boltzmann gas:}&{}&{}&{} \\
\m{non--relativistic~} (\beta\gg 1) & {\ts{5\over6}}\beta^{-2} p\tau_0
& {\ts{5\over2}}\beta^{-1}T^{-1}p\tau_1 & p\tau_2       \\
{}&{}&{}&{}
\\ \hline
\end{array}
$$

\[ \]
\[ \]
In the table, $\beta$ is given in standard units by
$$
\beta={mc^2\over kT}
$$
where $m$ is the mass of the matter particles (usually electrons);
$r_0$ is the radiation constant for photons, and
${7\over8}$ times the radiation constant for massless
neutrinos; $\Gamma$ is effectively
the deviation of $p/\rho$
from its pure--radiation value:
\be
\Gamma={\ts{1\over3}}-\left({\p p\over\p\rho}\right)_n=
{\ts{1\over3}}-{(\p p/\p T)_n \over (\p\rho/\p T)_n}
\l{6.}\ee
where $p,\rho$ refer to the pressure and energy density of the
radiation/ matter mixture as a whole. For example, when the matter
is non--relativistic, (\r{46}) and (\r{46c}) show that in
standard units
\be
p\approx nkT+{\ts{1\over3}}r_0T^4\,,\quad\quad\rho\approx
mc^2n+{\ts{3\over2}}nkT+r_0T^4
\l{7.}\ee
where $n$ is the number density of matter.

Note that for both the radiative fluid and the
Maxwell--Boltzmann gas, the bulk viscosity
tends to zero in
the ultra--relativistic and non--relativistic
limits. Bulk viscous effects are greatest in the mildly
relativistic intermediate regime, $\beta \approx 1$. This discussed
further in the next section.

The radiative fluid and Maxwell--Boltzmann gas are perhaps the
best motivated dissipative fluid models. However, their equations
of state and thermodynamic coefficients are very complicated,
and for the purposes of analytical rather than numerical
investigations, simplified equations are often assumed.
These are usually barotropic:
\be
p=p(\rho)\,,\quad T=T(\rho)\,,\quad \zeta=\zeta(\rho)\,,\quad
\lambda=\lambda(\rho)\,,\quad\eta=\eta(\rho)\,,\quad
\tau_A=\tau_A(\rho)
\l{9.}\ee
However
these assumptions are subject to consistency
conditions (as shown earlier in the case of $p$ and $T$),
and may correspond to unphysical behaviour. Whenever
such assumptions are made in a model, the consequences should be
carefully checked. An example is given in the next section.

\section{Causal Bulk Viscosity in Cosmology}

I will use the simplest case of scalar dissipation due to bulk
viscosity in order to illustrate some of the issues that arise
in modelling cosmological dissipation via Israel--Stewart theory.
Furthermore, this case covers the standard cosmological models.
If one assumes that the universe is exactly
isotropic and homogeneous -- i.e.
an FRW universe (\r{50}) -- then the symmetries show that only
scalar dissipation is possible -- i.e. $q_\a=0=\pi_{\a\b}$.
In this event, the no--coupling assumption (\r{17'}) is
automatically fulfilled.

Bulk viscosity arises typically in mixtures -- either of different
species, as in a radiative fluid,
or of the same species but with different energies, as in a
Maxwell--Boltzmann gas. Physically, we can think of bulk viscosity
as the internal `friction' that sets in due to the different cooling
rates in the expanding mixture. The dissipation due to bulk
viscosity converts kinetic energy of the particles into heat,
and thus we expect it to reduce the effective pressure in an
expanding fluid -- i.e. we expect $\Pi\leq 0$ for $H\geq 0$. This
is consistent with $\dot{S}\geq0$ by (\r{12a'}):
\be
Tn\dot{S}=-3H\Pi
\l{18.}\ee

Any dissipation in an exact FRW universe is scalar, and therefore
may be modelled as a bulk viscosity within a thermodynamical
approach. As I have argued in the previous chapter, the
Israel--Stewart thermodynamics is causal and stable under a wide
range of conditions, unlike the standard Eckart theory. Therefore,
in order to obtain the best thermo--hydrodynamic
model with the available physical
theories, one should use the causal Israel--Stewart theory of
bulk viscosity.

Writing out the full Israel--Stewart transport equation
(\r{19'}) (using $\tau\equiv\tau_0$), we get
\be
\tau\dot{\Pi}+\Pi=-3\zeta H-{\ts{1\over2}}\tau\Pi\left[3H+
{\dot{\tau}\over\tau}-{\dot{\zeta}\over\zeta}-{\dot{T}\over T}\right]
\l{12.}\ee

A natural question is -- what are the conditions under which
the truncated form (\r{22'}) is a reasonable approximation
of the full Israel--Stewart
transport equation? It follows from (\r{12.}) that if
\be
{T\over a^3H}\left|\Pi\left({a^3\tau\over\zeta T}\right)^{\ds}\right|
\ll 1
\l{13.}\ee
holds, then the additional terms in (\r{12.}) are negligible
in comparison with $3\zeta H$. The condition (\r{13.}) is clearly
very sensitive to the
particular forms of the functions $p(n,\rho)$, $\zeta(n,\rho)$
and $\tau(n,\rho)$. The temperature is determined on the basis
of these particular forms by the Gibbs integrability
condition (\r{30'}) and the evolution equation (\r{31'})\footnote{or
equivalently by (\r{32'}) and (\r{33'})}:
\be
{\dot{T}\over T}=-3H\left[\left({\p p\over\p\rho}\right)_n+
{\Pi\over T}\left({\p T\over\p\rho}\right)_n\right]
\l{19.}\ee
The second term on the right shows that bulk stress tends to
counteract the cooling due to expansion.

For simplicity, suppose that the pressure and temperature
are barotropic, with $p$ linear:
\be
p=(\gamma-1)\rho
\l{16.}\ee
This pressure equation is not unreasonable if the
local equilibrium state is radiation or cold matter.
Since the temperature is also barotropic,
it then
follows from the Gibbs integrability condition (\r{30'}) that
as in the perfect fluid case, $T$ must have the power--law
form (\r{46j}):
\be
T\propto \rho^{(\gamma-1)/\gamma}
\l{14.}\ee
Thus there is no freedom to choose the form of $T(\rho)$ -- it is
a power--law, with index
fixed by $\gamma$. With these forms of $p(\rho)$ and
$T(\rho)$, we can see that the temperature evolution equation
(\r{19.}) is identically satisfied by virtue of the
energy conservation equation (\r{3'}):
\be
\dot{\rho}+3H(\rho+p+\Pi)=0
\l{15.}\ee

A simple relation between $\tau$ and $\zeta$ is found
as follows. It is shown in the appendix to this chapter that
\be
{\zeta\over(\rho+p)\tau}=c_b^2
\l{11.}\ee
where $c_b$ is the speed of bulk viscous perturbations -- i.e. the
non--adiabatic contribution to the speed of sound $v$ in a dissipative 
fluid without heat flux or shear viscosity. The dissipative speed
of sound is given by
\be
v^2=c_s^2+c_b^2 \leq1
\l{10.}\ee
where $c_s$ is the adiabatic contribution (\r{39b}), and the
limit ensures causality. When (\r{16.}) holds, $c_s^2=\gamma-1$, so
that
$$
c_b^2\leq 2-\gamma
$$
We will assume that $c_b$ is constant, like $c_s$.

Putting together the thermodynamic relationships (\r{16.}),
(\r{14.}) and (\r{11.}), the full transport equation (\r{12.})
becomes
\be
\tau_*\dot{\Pi}+\Pi=-3\zeta_* H\left[1+{1\over\gamma c_b^2}
\left({\Pi\over\rho}\right)^2\right]
\l{17.}\ee
where the effective relaxation time and bulk viscosity are
\be
\tau_*={\tau\over 1+3\gamma\tau H}\,,\quad\zeta_*={\zeta\over
1+3\gamma\tau H}=c_b^2\gamma\rho\tau_*
\l{20.}\ee
Now the near--equilibrium condition (\r{26'}) with (\r{16.})
implies
$$
|\Pi|\ll\rho
$$
and shows that the second
term in square brackets in (\r{17.}) is negligible. Thus the full
equation leads to a truncated equation with {\em reduced relaxation
time and reduced bulk viscosity:}
\be
\tau_*\dot{\Pi}+\Pi=-3\zeta_* H
\l{23.}\ee
The amount of reduction depends
on the size of $\tau$ relative to $H$. If $\tau$ is of the order
of the mean interaction time, then the hydrodynamical description
requires $\tau H<1$. If $\tau H\ll 1$, then $\tau_*\approx\tau$
and $\zeta_*\approx\zeta$. But if $\tau H$ is close to 1, the
reduction could be significant.

Although this reduction is based on the simplified thermodynamical
relations assumed above, it indicates that the validity of
the truncated Israel--Stewart equation can impose significant
conditions.
More realistic thermodynamical relations will require numerical
calculations. In the case of
a Maxwell--Boltzmann gas, such calculations show that the
behaviour of the truncated and full theories can be very
different. The conclusion seems to be that the full theory should
be used, unless one is able to derive explicitly -- and satisfy -- the
conditions under which the truncated version is adequate.

Assuming that the FRW universe is flat, the Friedmann equation
(\r{49}) is
\be
\rho=3H^2
\l{21.}\ee
By (\r{15.}) and (\r{21.}) we get
\be
\Pi=-2\dot{H}-3\gamma H^2
\l{22.}\ee
and together with (\r{23.}) and (\r{20.}), this leads to the
evolution equation for $H$:
\be
\ddot{H}+(6\gamma+ N)H\dot{H}+{\ts{3\over2}}\gamma\left[
3(\gamma-c_b^2)+N\right]H^3=0
\l{24.}\ee
where
\be
N=(\tau H)^{-1}
\l{25.}\ee
is of the order of the number of interactions in an expansion time.
Intuitively, when $N\gg 1$, the fluid is almost perfect, while
when $N$ is close to 1, the dissipative effects are significant.
This is confirmed by (\r{24.}). For $N\gg 1$, the equation reduces
to
$$
\dot{H}+{\ts{3\over2}}\gamma H^2\approx 0
$$
with the well--known perfect fluid solution:
$$
H\approx{2\over 3\gamma (t-t_0)}
$$
On the other hand, for $N$ close to 1, the second derivative
in (\r{24.}) cannot be neglected, and the solutions will show a range
of behaviour very different from the perfect fluid -- and the
standard Eckart -- solutions. (Note that the Eckart limit $\tau
\rightarrow 0$ is $c_b\rightarrow\infty$ by (\r{11.}); the causality
condition (\r{10.}) does not hold.)

Of course, a complete model requires the specification of $N$.
Consider the ultra--relativistic fluid of the early universe,
with a particle species whose growing mean free path is giving rise to
dissipation, such as the neutrino. Suppose that $\tau\approx t_c$,
where $t_c$ is the mean interaction time, and that the interaction
cross--section is proportional to $T^2$, like the neutrino's.
Then by (\r{4.}) we get
\be
N=\left({T\over T_*}\right)^3=\left({H\over H_*}\right)^{3/2}
\l{26.}\ee
For $T\gg T_*$, we have $N\gg 1$, and dissipation is negligible. But
for $T$ close to $T_*$, dissipation effects become significant. The
evolution equation (\r{24.}) becomes
\be
\ddot{H}+\left[8+ \left({H\over H_*}\right)^{3/2}\right]
H\dot{H}+2\left[
4-3c_b^2+\left({H\over H_*}\right)^{3/2}\right]H^3=0
\l{27.}\ee
One could try to solve this equation perturbatively, by the ansatz
$$
H={1\over 2(t-t_0)}+\ep H_1+O(\ep^2)
$$
\[ \]

I will briefly discuss the question of bulk viscous inflation.
Suppose dissipation in the cosmic fluid produced sufficiently
large bulk viscous stress to drive the effective pressure
negative and thus initiate inflationary expansion.
By (\r{52}), using the effective pressure, the condition for
inflationary expansion is
\be
-\Pi>p+{\ts{1\over3}}\rho
\l{28.}\ee
For a fluid, this violates the near--equilibrium
condition
$$
|\Pi|\ll p
$$
Thus viscous fluid inflation, if it were physically possible, would
involve non--linear thermodynamics, far from equilibrium. The
Israel--Stewart theory, as well as other versions of extended 
thermodynamics and
also Eckart's standard thermodynamics, 
are all based on near--equilibrium
conditions, and cannot be applied to inflationary expansion --
unless one makes the drastic assumption that the linear theory
applies in the strongly non--linear regime.

Furthermore, there are serious physical problems with
hydrodynamic inflation (without particle production\footnote{See 
\cite{z}, \cite{gl2} for particle production models.}). 
The point is that under conditions of
super--rapid expansion -- i.e. very small expansion
time -- the hydrodynamic regime requires even smaller
interaction time. It is hard to see how the fluid interaction
rate could increase to stay above the expansion rate under
conditions where fluid particles are expanding apart from each
other extremely rapidly.

For a satisfactory model of bulk viscous inflation, one needs:
(a)~a non--linear generalisation of the Israel--Stewart transport
equation (\r{12.});\footnote{One possible generalisation is
developed in \cite{mm}.} (b)~a consistent model of fluid behaviour
under super--rapid expansion and strongly non--linear conditions.

On the other hand,
the reheating period at the end of inflation
can be modelled by near--equilibrium theory, and
the expansion rate is no longer inflationary. However, a
thermodynamic model needs to incorporate particle 
production.\footnote{See \cite{zpmc}.}
\[ \]

Finally, for those who like to analyse and solve differential 
equations\footnote{See also \cite{cj2} -- \cite{cjm}.}
more than they like physical analysis, I
will give the evolution equation of $H$ with mathematically
more general (but physically no more satisfactory) thermodynamic
equations of state. Suppose
$p$ and $T$ are given as above, by (\r{16.}) and (\r{14.}),
but instead
of the relation (\r{11.}), with constant $c_b$,
linking $\tau$ and $\zeta$, we assume
the barotropic forms
\be
\zeta\propto\rho^r\,,\quad\quad\tau\propto\rho^q
\l{29.}\ee
where $r$ and $q$ are constants.

Then with (\r{29.}), the (non--truncated)
evolution equation (\r{12.}) becomes
\bea
&&\ddot{H}+3\left[1+{\ts{1\over2}}(1+q-r)\gamma \right]H\dot{H}
+\alpha_1 H^{-2q}\dot{H}+\left(q-r-1+\gamma^{-1}\right)H^{-1}
\dot{H}^2  \nonumber\\
{}&&+{\ts{9\over4}}\gamma H^3+{\ts{3\over2}}\gamma\alpha_1
H^{2(1-q)}+{\ts{3\over2}}\alpha_2 H^{2r-2q+1} =0
\l{30.}\eea
where $\alpha_1$ and $\alpha_2$ are constants. One can find
special exact solutions, including exponential and power--law 
inflation, and perform a qualitative dynamical analysis of (\r{30.}),
or of similar equations arising from different forms for
the equations of state and thermodynamic coefficients.

\newpage

\section{Appendix: Bulk Viscous Perturbations}

A comprehensive analysis of the causality and stability properties
of the full Israel--Stewart theory has been performed by 
Hiscock and Lindblom \cite{hl}. They consider
general perturbations -- i.e. $\Pi, q_\a, \pi_{\a\b}$ all nonzero --
about a (global) equilibrium in flat spacetime, but the results
are valid in cosmology for short wavelength perturbations. In this
appendix, I will extract from their complicated
general results the special case
of scalar perturbations (only $\Pi\neq0$), when remarkably simple
expressions can be obtained.

The characteristic velocities for general dissipative perturbations
are given by equations (110) -- (128) in \cite{hl}. The purely
bulk viscous case is 
\be
\alpha_0=0=\alpha_1\,;\quad {1\over\beta_1}\,,~{1\over\beta_2}
~~\rightarrow~~0\,;\quad\beta_0\equiv {\tau\over\zeta}
\l{a1}\ee
(See (\r{15'}) and (\r{19'}) -- (\r{21a'}).)

Equation (127) of \cite{hl} gives the speed of the
propagating transverse modes: 
$$
v_T^2={(\rho+p)\alpha_1^2+2\alpha_1+\beta_1 \over
2\beta_2\left[\beta_1(\rho+p)-1\right]}~~\rightarrow~~0
$$
on using (\r{a1}). This is as expected for scalar sound--wave
perturbations. Equation (128) governing the speed $v=v_L$ of
propagating longitudinal modes becomes, on dividing by 
$\beta_0\beta_2$ and setting $\alpha_0=0=\alpha_1$:
\bea
\left[\beta_1(\rho+p)-1\right]v^4+\left[{2n\over T}\left({\p T\over
\p n}\right)_{\!S}-{(\rho+p)\over nT^2}\left({\p T\over\p S}\right)
_{\!n}-\beta_1\left\{(\rho+p)\left({\p p\over\p\rho}\right)_{\!S}
+{1\over\beta_0}\right\}\right]v^2 && \nonumber\\
{}+{1\over nT^2}\left({\p T\over\p S}\right)_{\!n}\left[(\rho+p)
\left({\p p\over\p \rho}\right)_{\!S}+{1\over\beta_0}\right]
-\left[{n\over T}\left({\p T\over\p n}\right)_{\!S}\right]^2=0 &&
\nonumber
\eea
Dividing by $\beta_1$ and taking the limit $\beta_1\rightarrow\infty$,
this gives
\be
v^2=\left({\p p\over\p\rho}\right)_{\!S}+{1\over(\rho+p)\beta_0}
\l{a2}\ee
The first term on the right is the adiabatic contribution $c_s^2$
to $v^2$, and the second term is the dissipative contribution $c_b^2$,
as in (\r{11.}).

It is also shown in \cite{hl} (pp 478--480)
that causality and stability require 
$$
\Omega_3(\lambda)\equiv(\rho+p)\left\{1-\lambda^2\left[\left({\p p
\over\p\rho}\right)_{\!S}+{1\over(\rho+p)\beta_0}\right]\right\}\geq0
$$
for all $\lambda$ such that $0\leq\lambda\leq1$. This condition
is shown to hold for all $\lambda$ if it holds for $\lambda=1$,
leading to the requirement 
\be
c_b^2\equiv{\zeta\over(\rho+p)\tau}\leq1-c_s^2
\l{a3}\ee
i.e. $v^2\leq1$, as expected. This establishes (\r{10.}).

These results refine and correct
the widely--quoted statement in \cite{bnk}
that $\zeta/\rho\tau=1$ is required by causality.

\vfill
\noindent
{\bf Acknowledgements}
\[ \]
Thanks to Sunil Maharaj for organising the Workshop so well and for
his wonderful hospitality. The participants at the Workshop helped
improve these notes by their questions and comments.
I was supported by a Hanno Rund Research Fellowship.
I have had many useful and inspiring discussions with
Diego Pavon, Winfried Zimdahl, David Jou, Josep Triginer, David 
Matravers and others.

\end{document}